\documentclass{ws-p8-50x6-00}

\begin{document}

\title{Berry Phase and the Symmetry of the Vibronic Ground State in
Dynamical Jahn-Teller Systems}

\author{Nicola Manini and Paolo De Los Rios}

\address{
European Synchrotron Radiation Facility,
B.P. 220, F-38043 Grenoble C\'edex, France Institut de Physique Th\'eorique,
Universit\'e de Fribourg, 1700-CH Fribourg, Switzerland }

\maketitle

\abstracts{
Due to the frequent presence of a Berry phase, in most cases of dynamical
Jahn-Teller systems the symmetry of the ground state is the same as that of
the electronic state.  However, the $H \otimes h$ icosahedral case,
relevant for the physics of fullerene ions, provides a first example of
linear coupling leading, at strong coupling, to a change in symmetry of the
ground state to a totally symmetric nondegenerate state.  We generalize
this observation and show through detailed examples that the absence of a
Berry phase can, but does not necessarily, lead to a nondegenerate ground
state.}

The traditional field of degenerate electron-lattice interactions
(Jahn-Teller effect) in molecules and impurity centers in
solids\cite{Englman,Bersuker} has attracted new interest in recent years,
excited by the realization of new systems which call for a revision of a
number of commonly accepted beliefs.  A whole range of icosahedral
molecular systems including C$_{60}$ ions and some higher fullerenes,
thanks to the rich structure of the symmetry group, are characterized by up
to fivefold-degenerate representations of the electronic and vibrational
states of the isolated molecule/ion.  Novel Jahn-Teller (JT) systems have
therefore been considered theoretically,\cite{Bersuker,Ihm,AMT} disclosing
intriguing features,\cite{AMT,Mead,Wilczek,Delos96,Moate96} often related
to the r\^ole of a Berry phase\cite{Berry} in the coupled dynamics.

As it is well known, the molecular symmetry, reduced by the JT distortion
with the splitting of the electronic-state degeneracy, is restored when the
coherent tunneling between equivalent distortions is considered, in the
dynamical Jahn-Teller (DJT) effect.  In this context it was commonly
accepted an empirical ``symmetry conservation rule'', sometimes referred to
as ``Ham's theorem'', stating that the symmetry of the vibronic DJT ground
state, at all coupling strengths, remains the same as that of the
electronic multiplet prior to coupling:\cite{Bersuker} all linear JT
systems known till a few years ago, for single-electron occupancy,
systematically satisfy this empiric rule.  It was understood recently that
this phenomenon, not automatically implied by the DJT physics, is in
reality a fingerprint of a Berry phase\cite{Berry} in the entangled
electronic-phononic dynamics.\cite{AMT,Delos96,ob96} Consequently, this
geometrical phase appeared as a universal feature of the DJT systems.

In this context, it came unexpected the discovery of the first linear JT
system showing a {\em nondegenerate ground state} in the strong-coupling
limit.\cite{Delos96,Moate96} This result was demonstrated for the model
that in spherical symmetry is indicated as ${\cal D}^{(2)}\otimes d^{(2)}$,
where electrons of angular momentum $L=2$ interact with vibrations also
belonging to an $l=2$ representation.  This system is relevant to the
physics of fullerene ions C$_{60}^+$, where the 5-fold degenerate
electronic state has $H_u$ icosahedral label and the quadrupolar
distortions correspond to some of the $h_g$ modes.\cite{Delos96} It has
been shown by different methods and independent groups that, for increasing
coupling, a nondegenerate state in the vibronic spectrum moves down, to
cross the 5-fold ground state at some finite value of the coupling
parameter, thus becoming the ground state at strong
coupling~.\cite{Delos96,Moate96} This phenomenon is related to the absence
of a Berry phase entanglement in the coupled dynamics.\cite{Delos96}

The r\^ole generally attributed to the Berry phase is therefore to
guarantee a ``symmetry conservation rule'' for the ground state from weak
to strong coupling of DJT systems.  The absence of this geometrical phase
allows the strong-coupling ground state to become the ``natural''
nondegenerate totally symmetrical representation that a naive picture,
ignoring this geometrical phase, would predict in all cases.  In this work
we reconsider in detail the connection between the symmetry/degeneracy of
the vibronic ground state of a large class of DJT systems, and the
presence/absence of a Berry phase in the coupled dynamics, finding that the
relation sketched above does not apply automatically to all cases.

In the general formalism of the JT effect, a degenerate electronic state
corresponding to a representation $\Gamma$ of the symmetry group $\cal G$
of the molecule can interact with the vibrational modes corresponding to
representations $\{\Lambda\}$ contained in the symmetric part of the direct
product $\Gamma \otimes \Gamma$ (excluding the identical representation
which is trivial).  In the case where exactly one mode of each symmetry
label $\Lambda$, of frequency $\omega_{\Lambda}$ and coordinates
$q_{\Lambda i}$, interacts linearly with strength $g_{\Lambda}$ with the
$|\Gamma|$-fold degenerate electronic level (with a fermion operator
$c_{\Gamma k}$), the Hamiltonian may be written:
\begin{equation}
H = \frac{1}{2} \sum_{\Lambda} \hbar \omega_{\Lambda}
\sum_{i=1}^{|\Lambda|} (p_{\Lambda i}^2 + q_{\Lambda i}^2) + H_{\rm e-v}\;,
\label{hamiltonian}
\end{equation} with
\begin{equation}
H_{\rm e-v} = \frac 12 \sum_{\Lambda} g_{\Lambda} \hbar
\omega_{\Lambda}
\sum_{i=1}^{|\Lambda|} \sum_{j,k=1}^{|\Gamma|}
q_{\Lambda i} c^\dagger_{\Gamma j} c_{\Gamma k} \langle\Lambda i | \Gamma j
\Gamma k\rangle\;,
\label{interaction hamiltonian}
\end{equation}
where $\langle\Lambda i | \Gamma j \Gamma k\rangle$ are the Clebsch-Gordan
coefficients for the group $\cal G$.\cite{butler81} In
Eq.~(\ref{hamiltonian}) we choose the real representation for the
vibrational degrees of freedom, and a second-quantized notation for the
electrons.

In the general case of arbitrary frequencies $\omega_\Lambda $ and
couplings $g_\Lambda$, the point group symmetry $\cal G$ is reflected in
the JTM, constituted of isolated minima, separated by saddle points.
However, the continuous JTM of the special equal-coupling equal-frequencies
case is invariant for transformations in the group $SO(|\Gamma|)$ of the
electronic manifold.  Indeed, the whole problem reduces to a single-mode JT
coupling between two representations of that group of
$|\Gamma|$-dimensional rotations.\cite{Pooler,Judd84} In such a case, it is
well known\cite{Ceulemans} that the set of minima of the Born-Oppenheimer
(BO) potential, corresponding to the most energetically-favorable classical
distortions, constitute a continuous manifold, referred to as Jahn-Teller
manifold (JTM).  The JT coupling induces an adiabatic mapping of the
vibrational space into the electronic space. Here we only sketch this
mapping, which is described in greater detail elsewhere.\cite{noberry}

In the traditional BO scheme there are assumed much larger separations
between consecutive electronic levels than the typical vibrational energies
$\hbar \omega$.  In a JT problem, each electronic eigenvector $|\psi_\xi
\rangle$ of the coupling matrix (\ref{interaction hamiltonian}), of
eigenvalue $\lambda_{\xi}$, generates a BO potential sheet $V_{\xi}(\vec
q)$.  At strong coupling $g$, the separation of the potential sheets
becomes so large that the adiabatic motion can be safely assumed to always
follow the lowest BO potential sheet, while virtual electronic excitations
may be treated as a small correction.

\begin{figure}
\epsfxsize 10.0cm
\epsfbox{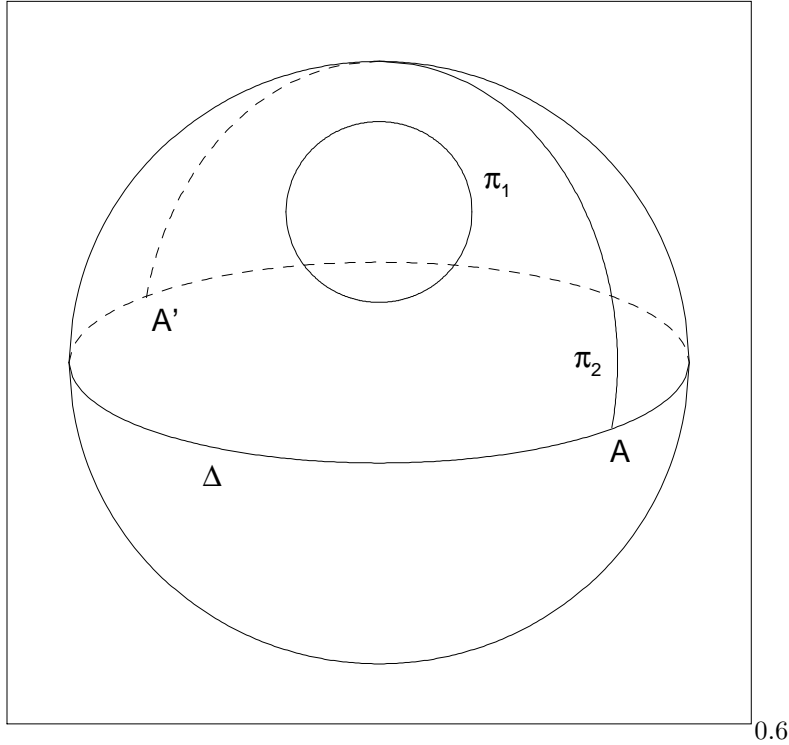}{0.6}
\caption{ A sketch of the electronic sphere.  The picture individuates the
two classes of paths mapping onto closed loops in the JTM: paths of the
type $\pi_1$ may be contracted continuously to a single point, while those
of type $\pi_2$ involve a sign change (from A to A') of the electronic
state (a Berry phase).\label{pathcl:fig}}
\end{figure}

On the other side, due to time-reversal invariance of $H$, the space of all
possible (normalized) electronic eigenstates can be represented by an
(hyper-)sphere in the $|\Gamma|$-dimensional real space (see Fig.\
\ref{pathcl:fig}).  The BO dynamics realizes an adiabatic mapping of the
vibrational space into this electronic sphere:\cite{Ceulemans} every point
$\vec{q}$ on the JTM (in the vibrational space) is associated to the
electronic wave function $|\psi_{\rm min} (\vec{q})\rangle$, corresponding
to the lowest eigenvalue $\lambda_{\rm min}$ of the interaction matrix.

\noindent

This adiabatic mapping is two-valued, since opposite points $\pm |\psi_{\rm
min}(\vec{q})\rangle$ on the electronic sphere give the same JT
stabilization energy, thus corresponding to the same optimal distortion on
the JTM.  This identification of the antipodal points through the mapping
is the mechanism allowing the JTM to have a different (topology) with
respect to the electronic sphere.  The latter is always simply connected,
i.e.\ any closed path on it can be smoothly contracted to a single point.
The JTM, instead, may well be multiply connected, i.e.\ it can have
intrinsic ``holes'' in its topology, related to the nontrivial class of
those loops mapped on a path going from a point to its antipode on the
electronic sphere, such as $\pi_2$ in Fig.\ \ref{pathcl:fig}.  This
electronic sign change is a case of Berry phase.\cite{Berry}

This geometric phase acts as a boundary condition for the quantization of
the vibrational motion.  As a consequence, the motion on the JTM is
constrained by special selection rules.  For example the JTM of the simple
$E\otimes e$ system is a circle: the low-energy vibronic spectrum is indeed
a $j^2$ spectrum as for a circular rotor, but the Berry phase implies
$j=\pm\frac 12,\pm\frac 32,...$, instead of $j=0,\pm 1,\pm 2,...$ as for an
ordinary quantum rotor.\cite{Bersuker,Koizumi} Similarly, the JTM of the
$T\otimes h$ (i.e.\ ${\cal D}^{(1)} \otimes d^{(2)}$, in the spherical
language) is equivalent to a sphere,\cite{AMT,ob71} but out of all the
states, labeled by $J,M$, of a particle on a sphere, the Berry phase
retains only the odd-$J$ ones.\cite{Bersuker,AMT,ob71} Note in particular
that in these examples the presence of a Berry phase rules out the
``natural'' nondegenerate ground state, and enforces, to the
strong-coupling DJT ground state, the same original symmetry $\Gamma$ of
the degenerate electronic state.

The Berry phase, though {\em not} automatically implied by linear JT
Hamiltonians (\ref{hamiltonian}), is indeed a very common feature.  The
double-valuedness of the adiabatic mapping described above is unavoidable.
For the Berry-phase--free cases, the mechanism leading to equivalence of
the paths in the class 1 and 2 needs to coexist with it.  As demonstrated
in earlier work,\cite{Delos96,noberry,Paris97} the solution of the riddle
is provided by a point $\vec q_d$ {\em on the JTM} where the mapping is
degenerate, i.e.\ it links $\vec q_d$ not just to a pair of opposite points
$\pm |\psi_{\rm min}(\vec q_d)\rangle$ on the electronic sphere, but to the
whole circle (such as, for example, $\Delta$ in Fig.\ \ref{pathcl:fig}) of
linear combinations $\cos \theta ~|\psi_1(\vec q_d)\rangle+ \sin \theta
 ~|\psi_2(\vec q_d)\rangle $ of two degenerate orthogonal electronic
eigenstates.  Where such a point is present, it allows to deform smoothly
any loop of class 2 on the JTM, until its image on the electronic sphere
becomes half this circle, thus shrinks to the single point $\vec q_d$.  All
loops are therefore contractable, thus equivalent to one another and,
therefore, the JTM is simply connected.  No Berry phase is possible in such
a case.\cite{noberry,Paris97}

Such a tangency point is the origin of the inversion of the low-lying
levels in the $H\otimes h$ JT problem,\cite{Delos96,noberry} leading to a
nondegenerate ground state at strong coupling.  Similar tangential points
were demonstrated\cite{noberry,Paris97} in other spherically symmetric
linear models, the ${\cal D}^{(L)}\otimes d^{(L)}$, with $L=2,4,6,...$.
All these systems are therefore Berry-phase free, with, in particular, a
strong-coupling non-degenerate vibronic ground state.  A numerical test
confirms this result in the ${\cal D}^{(4)}\otimes d^{(4)}$ case.  On the
contrary, these tangencies are absent in most DJT cases ($E\otimes e$,
$T\otimes h$, ${\cal D}^{(2)}\otimes d^{(4)}$, ...), whence the Berry
phase, whence the degenerate ground state at strong coupling.

The systems ${\cal D}^{(L)}\otimes d^{(l)}$, with $L>l$, are remarkable in
having a tangency point, thus no Berry phase as the preceding example, but
{\em no symmetry change} of the ground state, which remains degenerate to
all couplings.\cite{noberry} This case should be kept as a warning against
the simplistic equation: absence of Berry phase = nondegenerate
strong-coupling ground state.

We move on now to the investigation of the relations between ground state
symmetry, Berry phases and tangencies of potential sheets in the more
general case of $H \otimes (2 \; h \oplus g)$ Jahn-Teller coupling in
icosahedral symmetry.  The two $h$ and the $g$ modes can be classified
according to their spherical parentage
\begin{eqnarray}
d^{(2)} & \rightarrow & h_{[2]} \nonumber \\
d^{(4)} & \rightarrow & h_{[4]} \oplus g \;.
\label{fragmentation}
\end{eqnarray}
The existence of two different couplings $g_{h_{[2]}}$ and $g_{h_{[4]}}$ to
modes $h$ reflects the fact that the icosahedral group is not simply
reducible:\cite{Hamermesh} two independent sets of Clebsch-Gordan
coefficients for the coupling of $h$ and $h$ to $h$ are
necessary.\cite{Moate96,butler81,IhCGcoef:note} In the special case when
$\omega_{h_{[4]}}=\omega_g = \omega_4$ and $g_{h_{[4]}} = g_g = g_4$ the
$SO(3)$ symmetry of the linear problem is restored, and it can be labeled
accordingly: ${\cal D}^{(2)}\otimes (d^{(2)}\oplus d^{(4)})$).  In the
limit $g_4 =0$ we recover the Berry-phase--free ${\cal D}^{(2)}\otimes
d^{(2)}$ model discussed above.

In the completely equal-coupling equal-frequencies limiting case $\omega_4
= \omega_2$ and $g_4 = g_2$, as anticipated above for the general case, the
symmetry rises further to $SO(5)$:\cite{Judd84} the model may be described
as $[1,0]~\otimes~[2,0]$ in the notation of $SO(5)$ representations.  For
the equal-coupling case, the presence of a Berry phase has been explicitly
demonstrated,\cite{Delos96} together with its consequences for the
selection rules on the levels: it favors in the low-energy end of the
spectrum $[k,0]$ levels with odd $k$.  In particular, it was verified that
the ground state remains $5$-fold degenerate ($[1,0]$ in $SO(5)$ notation,
i.e.\ ${\cal D}^{(2)}$ as a $SO(3)$ representation), and the first excited
is a $30$-fold degenerate $[3,0]$ level.

In the general case $g_2\neq g_4$, the symmetry reduces to $SO(3)$, thus
the large $SO(5)$ representations split into their spherical components.
In sweeping the value of $g_4$ from $g_2$ down to $0$, the system passes
smoothly from a regular Berry-phase--related degenerate ground state to the
${\cal D}^{(2)}\otimes d^{(2)}$ Berry-phase--free nondegenerate ground
state (for large enough $g_2$).  In this final situation, the degenerate
${\cal D}^{(2)}$ state takes the r\^ole of the lowest excited state,
separated by a finite energy gap from the nondegenerate ${\cal D}^{(0)}$
ground state.\cite{Delos96,Moate96} Thus, a level crossing takes place
between the low-lying levels, at some intermediate value of $g_4$: we can
define a crossover value $g_4^{\rm c}$ (dependent on $g_2$) for which the
ground-state symmetry changes.  At strong coupling, the energy gap
$E[L=2]-E[L=0] = c_2 /g_2^2 + O(g_2^{-4})$ for $g_4=0$ and $E[L=2]-E[L=0] =
-c_4 /g_4^2 + O(g_4^{-4})$ for $g_2=0$, where $c_2$ and $c_4$ are positive
constants depending on $\omega_2$ and $\omega_4$.  Thus, the crossover
curve $(g_2,g_4^{\rm c}\left(g_2\right))$ should get asymptotically close
to the straight line $g_4=g_2 \left(c_4 /c_2\right)^{1/2}$ in the plane of
the coupling parameters.

\begin{figure}[h] 
\epsfxsize 10.0cm 
\epsfbox{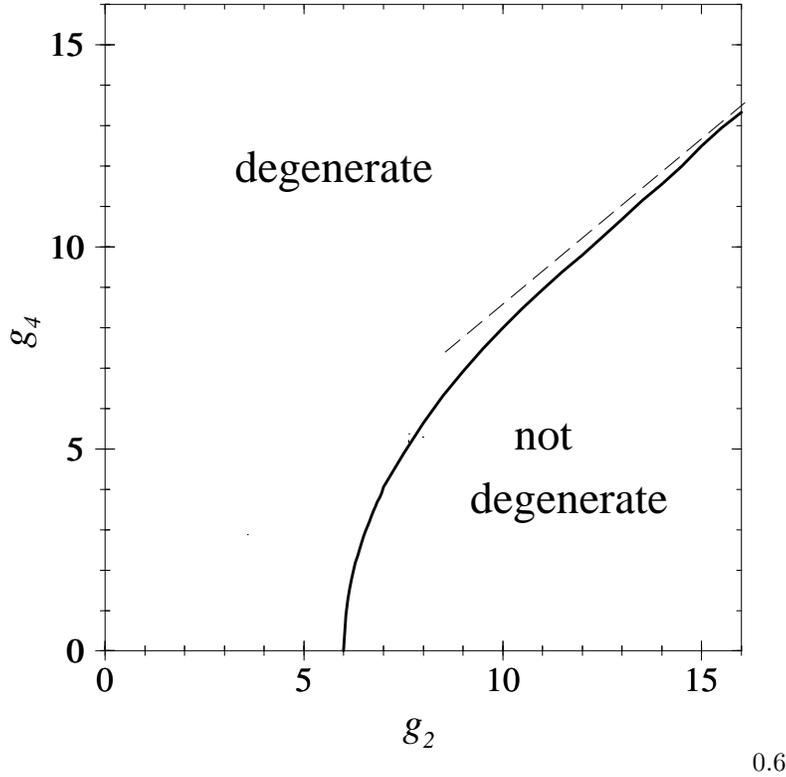}{0.6}
\caption{ The zero-temperature ``phase diagram'' of the ${\cal
D}^{(2)}\otimes (d^{(2)}\oplus d^{(4)})$ JT system in the space of the
coupling parameters $g_2$ and $g_4$, for fixed frequencies $\omega_2$ and
$\omega_4$.  At the solid line $g_4=g_4^{\rm c}$ the $L=0$ and $L=2$ ground
states become (accidentally) degenerate.
\label{phased:fig}}
\end{figure}

These considerations, as well as some exact diagonalizations on a truncated
basis, permit to draw the qualitative zero-temperature ``phase diagram''
represented in Fig.~\ref{phased:fig}.  It strikes for containing a whole
region ($0<g_4<g_4^{\rm c} \left(g_2\right)$) where a nondegenerate $L=0$
ground state coexists with the presence of a Berry phase.  This example
should stand as a warning against the simplistic equation: Berry phase =
degenerate strong-coupling ground state of the same symmetry as the
non-interacting electronic state.  We conclude, accordingly, that the
presence of a Berry phase in many-mode DJT systems is {\em not a sufficient
condition} for the degeneracy of the ground state. Indeed, even if the
overall system has a phase entanglement, the absence of a Berry phase in
one of the single-mode couplings allows for a non-degenerate ground state
in some regions of the coupling-parameters space.

\noindent

At this point it is necessary to reconcile the gradual, smooth lowering of
the nondegenerate state as $g_4/g_2$ is reduced from equal coupling towards
zero, with the abrupt disappearance of the Berry phase (which is a
topological effect, intrinsically non-per\-tur\-ba\-tive) for $g_4 = 0$.
The origin of the nondegenerate state is to be traced back to the 30-fold
degenerate first-excited state ($[3,0]$ according to $SO(5)$) of the
equal-coupling ``hypersymmetrical'' spectrum which splits into its
$L=0,3,4,6$ components ($SO(3)$ representations) as soon as $g_4\neq g_2$.
In particular, this $L=0$ fragment is the lowest when $g_4/g_2<1$.  For
small enough $g_4 /g_2$, this nondegenerate state has the opportunity to
localize as much as possible in the potential well in the $d^{(2)}$ vibron
space (corresponding to the JTM in the space of $d^{(2)}$ vibrations),
eventually crossing down below the $L=2$ ground state, to become itself the
ground state.  Even in this region, however, the Berry-phase prescription
in the $SO(5)$ language is respected, since the $L=0$ state is indeed a
fragment of an odd ($[3,0]$) -- Berry-phase allowed -- level: the ground
state still fulfills the parity constraint imposed to the low-energy
$SO(5)$ representations by the Berry phase in the global space.

Note, incidentally, that, although the ${\cal D}^{(2)}\otimes d^{(2)}$
problem is only $SO(3)$-symmetric, its JTM has $SO(5)$ symmetry.
Therefore, in the limit of infinitely large $g_2$ and vanishing $g_4$,
where the motion is essentially restricted to the JTM in the $d^{(2)}$
vibration space, that same nondegenerate ground state may also be
classified as an $[0,0]$ state for the symmetry group of the JTM, where it
complies therefore with the absence of Berry phase.

As a first remark, we note that our treatment calls for a revision of the
customary association of Berry's phase to a breakdown of the BO
approximation.  Indeed, the geometrical phase originates at the conical
intersections of the lowest two BO sheets.  At strong coupling, such points
lie at high energy and the system explores them with extremely small
probability.  On the contrary, here we relate the {\em absence} of the
geometrical phase to tangential contacts of the adiabatic sheets, {\em on
the JTM}, thus affecting low-potential regions which the system occupies
currently.
Thus, in these systems, it is not the Berry phase which is
connected to a breakdown of the BO approximation, but its absence.

Our analysis considers for simplicity spherical DJT models: however, it can
be extended to molecular point groups.  For example, the Berry phase and
ground-state symmetry switch of the $H\otimes \left(h_{[2]} +
h_{[4]}\right)$ are completely analogous to those of ${\cal D}^{(2)}\otimes
(d^{(2)}\oplus d^{(4)})$ described above.\cite{Moate96}

Also, we assume a linear JT coupling scheme (Hamiltonian
(\ref{hamiltonian})), which is the less realistic, the stronger the JT
distortion.  The introduction of quadratic and higher-order couplings has
usually effects similar to those produced by unequal linear couplings
and/or frequencies in $T\otimes (e+h)$ in cubic symmetry,\cite{ob69} i.e.\
of ``warping'' the JTM, reducing its symmetry.  Yet, the connectedness
properties are topological properties, thus robust against warping, as long
as it can be treated as a perturbations.  To quote the simplest example,
the introduction of quadratic terms in the $e \otimes E$
Hamiltonian\cite{Ham87} does not substantially change the picture as far as
the Berry phase and the symmetry/degeneracy of the ground state are
concerned.  In fact, even at strong JT coupling, the tunneling among rather
deep isolated minima is affected by the electronic phase,\cite{Bersuker}
and, as a result, the lowest tunnel-split state retains the same symmetry
and degeneracy as in the purely linear-coupling case.  Of course, if the
quadratic and higher-order couplings dominate over the linear term, new
conical intersections may appear, thus affecting the Berry phase and,
consequently the ground-state symmetry.\cite{BersukerAErice}

In summary, the standard r\^ole of the Berry phase is to guarantee a
``symmetry conservation rule'' for the ground state from weak to strong
coupling of linear DJT systems. Here, we propose two counterexamples to
this simple pattern: (i) a whole family, of Berry-phase--free dynamical JT
systems with a {\em degenerate} ground state at all couplings, the ${\cal
D}^{(L)}\otimes d^{(l)}$ models, with $l<L$; and (ii) the case of many
modes coupled at the same time to an electronic state, some with a Berry
phase entanglement, and some without it, in the region where the coupling
to the seconds prevail, the strong-coupling ground state can switch to
nondegenerate, as we illustrate for ${\cal D}^{(2)}\otimes (d^{(2)}\oplus
d^{(4)})$.  This second point, in particular, for those cases, such as
positive fullerene ions, where Berry-phase--free modes are present,
underlines the relevance of the actual values of the coupling strengths
between degenerate electrons and vibrations, which only permit to make a
prevision about the actual symmetry of the vibronic ground state.  In this
perspective, the experimental or {\em ab-initio} determination of the
detailed values of such couplings is of the utmost importance for this
class of systems.  Finally, the r\^ole of the Berry phase being that of
ordering the strong-coupling spectrum, it is conceivable a system where the
geometric phase enforces a non-totally symmetrical vibronic state of {\em
symmetry other than $\Gamma$}, that of the original electronic state:
further investigation od the icosahedral JT zoology may find a realization
of this possibility.

\section*{Acknowledgement}

We thank Arnout Ceulemans, Brian R. Judd, Erio Tosatti, and Lu Yu for
useful discussions.


\begin{thebibliography}{10}
\bibitem{IhCGcoef:note}
{ In the literature, different choices of these Clebsch-Gordan coefficients are
  available, according to the combination of $h_{[2]}$ and $h_{[4]}$ being
  taken as basis set}.
\end{thebibliography}
\end{document}